\documentclass[longbibliography,10pt]{iopart}

\usepackage{lineno}
\usepackage{bm} 
\usepackage{graphicx} 
\usepackage{comment} 
\usepackage{textcomp} 
\usepackage{color,soul}
\usepackage{caption}

\usepackage[colorinlistoftodos]{todonotes}
\usepackage{siunitx}
\usepackage{enumitem}
\setlist{noitemsep,leftmargin=*,topsep=0pt,parsep=0pt}
\usepackage{flushend}
\usepackage{upgreek}

\usepackage{xr}
\makeatletter

\newcommand*{\addFileDependency}[1]{
\typeout{(#1)}
\@addtofilelist{#1}
\IfFileExists{#1}{}{\typeout{No file #1.}}
}\makeatother

\usepackage{xcolor} 
\definecolor{lightgray}{gray}{0.6}
\definecolor{medgray}{gray}{0.4}

\usepackage{hyperref}
\hypersetup{
colorlinks=true,
urlcolor= blue,
citecolor=blue,
linkcolor= blue,
}

\newif\ifptitle
\newif\ifpnumber
\newcounter{para}

\ptitletrue  
\pnumbertrue  

\begin{document}
\title{Temporal solitons in active optical resonators}

\author{D. Kazakov$^1$, F. Capasso$^1$ and
M. Piccardo$^{1,2,3}$
}

\address{$^1$ Harvard John A. Paulson School of Engineering and Applied Sciences, Harvard University, Cambridge, MA 02138, USA}
\address{$^2$ Department of Physics, Instituto Superior Técnico, Universidade de Lisboa, Lisbon, Portugal}
\address{$^3$ Instituto de Engenharia de Sistemas e Computadores—Microsistemas e Nanotecnologias (INESC MN), Lisbon, Portugal}
\ead{marco.piccardo@tecnico.ulisboa.pt}

\begin{abstract}

Solitons, as coherent structures that maintain their shape while traveling at constant velocity, are ubiquitous across various branches of physics, from fluid dynamics to quantum fields. However, it is within the realm of optics where solitons have not only served as a primary testbed for understanding solitary wave phenomena but have also transitioned into applications ranging from telecommunications to metrology. In the optical domain, temporal solitons are localized light pulses, self-reinforcing via a delicate balance between nonlinearity and dispersion. Among the many systems hosting temporal solitons, active optical resonators stand out due to their inherent gain medium, enabling to actively sustain solitons. Unlike conventional modelocked lasers, active resonators offer a richer landscape for soliton dynamics through hybrid driving schemes, such as coupling to passive cavities or under external optical injection, affording them unparalleled control and versatility. We discuss key advantages of these systems, with a particular focus on quantum cascade lasers as a promising soliton technology within the class of active resonators. By exploring diverse architectures from traditional Fabry-Perot cavities to racetrack devices operated under external injection, we present the current state-of-the-art and future directions for soliton-based sources in the realm of semiconductor lasers and hybrid integrated photonic systems.

\end{abstract}

\maketitle

\tableofcontents

\pagebreak

\section{Introduction}

\subsection{From water waves to optical solitons}

\begin{figure}[!b]
    \centering
    \includegraphics[width=0.5\textwidth]{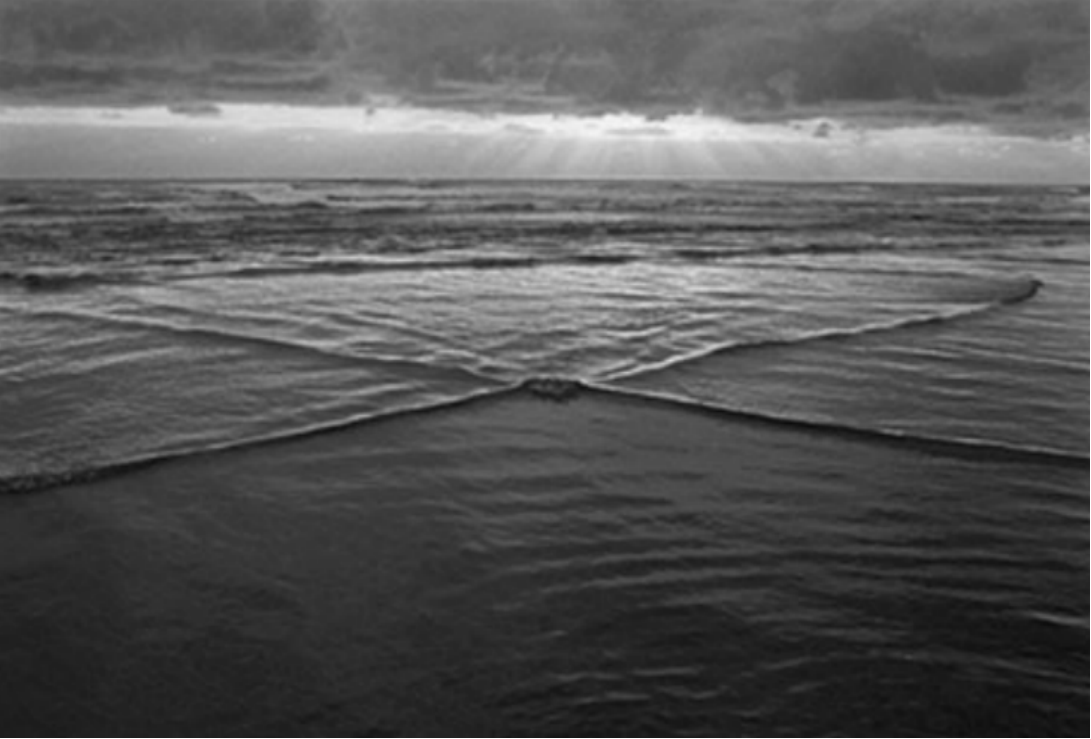}
    \captionsetup{width=0.5\textwidth}
    \caption{Line solitons, localized in one direction and infinitely extended in the other, observed on the Oregon coast. Reproduced from \cite{Klein2012}.}
    \label{fig:linesoliton}
\end{figure}

The phenomenon of solitons, first observed by John Scott Russell in 1834, has progressed from an historical observation to a fundamental concept in modern physics. Russell's initial encounter with a solitary wave in a shallow canal --- a wave maintaining its shape while traveling at a constant speed --- laid the groundwork for a deeper understanding of this phenomenon. Today, we can recognize the presence of solitons far beyond canals; they are discernible even in the vast expanses of nature, such as the line solitons admired along the ocean coast (Fig. \ref{fig:linesoliton}). Technological advancements in hydrodynamic systems have further sharpened our study of water solitons. No longer confined to natural settings, we can now routinely generate and observe solitons in laboratory tanks in controlled environments across various scales, from meters~\cite{Redor2019} to centimeters~\cite{Novkoski2022}, revealing a plethora of phenomena ranging from head-on collisions to the formation of soliton gases.

The universal appeal of solitons lies in their occurrence as nonlinear localized wave solutions in various celebrated equations in physics, including the Korteweg–de Vries (KdV) and the nonlinear Schrödinger  equations. Korteweg and de Vries discovered the one-soliton solution in 1895 in their shallow water wave theory, formalizing the empirical observations of Scott Russell \cite{korteweg1895xli}. The KdV equation was rediscovered much later, in 1955, in connection with one of the first dynamics simulations carried out on a computer, by Fermi, Pasta, Ulam and Tsingou (FPUT) who found that a chain of harmonic oscillators coupled with a nonlinearity cycled quasi-periodically \cite{fermi1955studies}. Understanding of the behavior of solutions was then advanced in 1965 by Zabusky and Kruskal \cite{zabusky1965interaction}, who rederived the KdV equation as the continuum limit of the FPU model, reporting numerical observations of KdV solitons and connecting these to the FPUT recurrence.

The unique shape of solitons, originating from a delicate balance between nonlinear and dispersive effects, allows for their occurrence in diverse media, from fluid dynamics and granular media such as sand dunes~\cite{Schwaemmle2003,ankiewicz2008dissipative,akhmediev1997nonlinear} to more complex systems like plant ecology, chemical reaction-diffusion systems, mechanical structures, microwave transmission lines, and Bose-Einstein condensates. In the realm of optics, the analogy with hydrodynamics has been central in advancing our understanding of nonlinear optics. Optical solitons, akin to their hydrodynamic counterparts, exhibit a wide range of fascinating properties, such as plasticity and multistability, but are also finding ways into practical applications. The trend towards miniaturization of optical sources has shifted soliton applications from the original concepts of all-optical memories and delay lines to contemporary uses in coherent communications, ranging, atomic clocks, dual-comb spectroscopy, and optomechanical cavities.

\subsection{Focus of this review}

While the study of optical solitons is vast and well-explored in several excellent books \cite{descalzi2011localized,Akhmediev2005,ackemann2009fundamentals,taylor1992optical}, our review seeks to focus on a specific and emerging sub-branch of this field: solitons in active optical resonators. The introduction of gain in these systems does not merely transform an input beam into a soliton; it enables the generation of solitons, presenting novel dynamics and applications. Active resonators do not necessarily have to be operated above the lasing threshold and can be operated in hybrid driving schemes, such as by injecting an external signal or being coupled to passive cavities, enriching the control possibilities and soliton behaviors of these systems. Our review will first revisit key general notions of optical solitons, then delve into the distinctions between passive and active systems, and ultimately focus on the emerging field of miniaturized soliton laser technology.

\section{Key notions in optical solitons}



\subsection{Modulation instability: From optical patterns to solitons}

\begin{figure}[!t]
    \centering
    \includegraphics[width=0.5\textwidth]{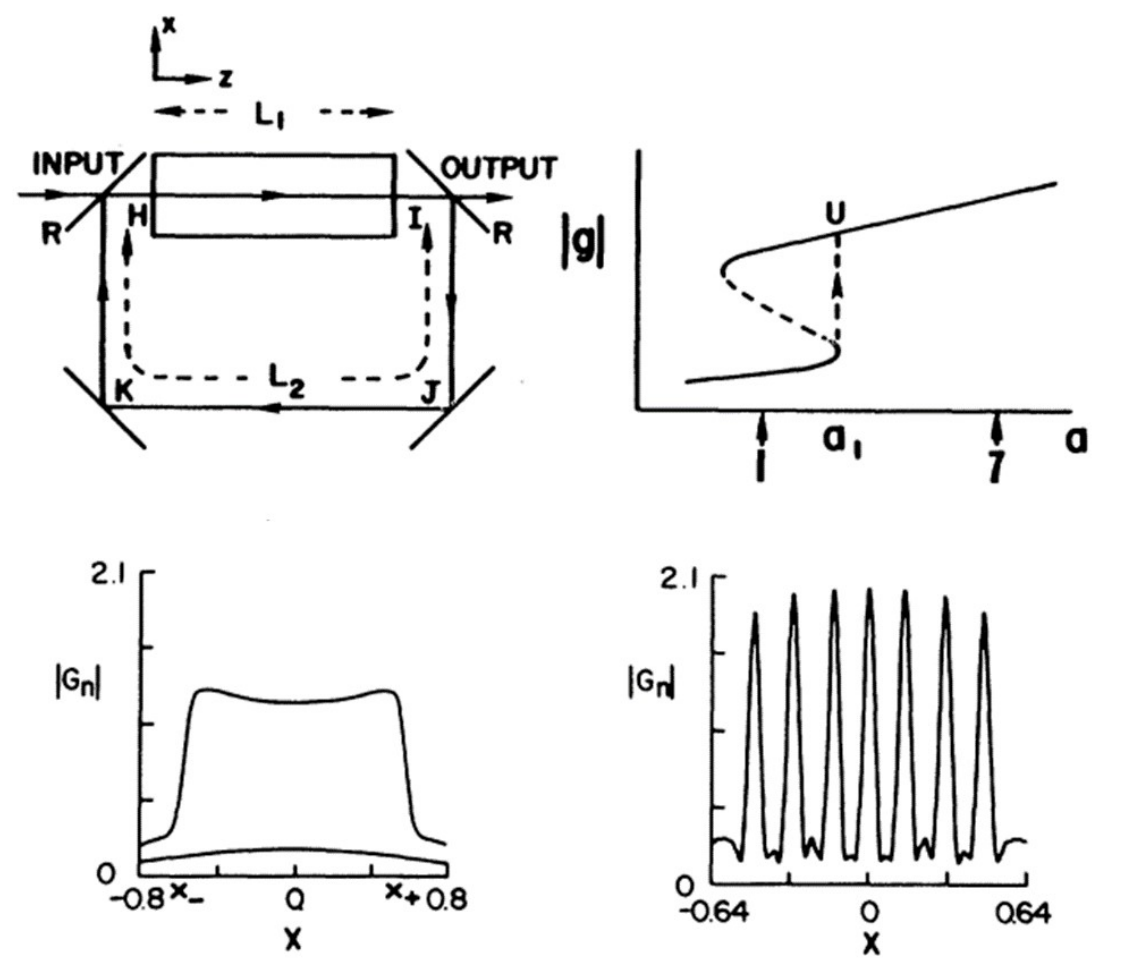}
    \captionsetup{width=0.5\textwidth}
    \caption{Simulation by Mc Laughlin and co-workers carried out in 1983 representing a historical landmark in the framework of cavity solitons. The system consists of a simple injected optical ring cavity containing a nonlinear saturable medium (top left) resulting in a bistable loop in the output vs. input field (top right). An input beam with a transverse Gaussian profile can ``switch-up'' its intensity thanks to the bistability (bottom left) and after 200 cavity roundtrips form a steady-state seven-solitary-wave train (bottom right). Reproduced from \cite{Laughlin1983}.}
    \label{fig:mclaughlin}
\end{figure}

A key milestone in the history of optical solitons is the simulation conducted by Mc Laughlin and co-workers~\cite{Laughlin1983}. We describe it here as this allows us also to highlight general notions of solitons emerging in an optical system. The simulation, based on the split-step fast Fourier transform (FFT) method, considered the injection of a Gaussian beam into a ring cavity hosting a Kerr medium, demonstrating a fundamental mechanism by which cavity solitons can arise (Fig. \ref{fig:mclaughlin}).

Increasing the power of the input beam, it was observed that the central part of the Gaussian beam underwent a ``switch-up'' in energy. This switch-up event---a central part of the beam suddenly intensifying---is a classic demonstration of optical bistability within a Kerr medium. The Kerr medium's refractive index changes with the light intensity, which leads to the bistable behavior where two distinct states can exist: one with lower transmission (``off'') and another with higher transmission (``on'').

After several roundtrips in the cavity, the subsequent development observed was modulation instability. This instability caused small perturbations on the otherwise uniform beam to grow, ultimately forming a pattern of localized peaks. These peaks were identified as cavity solitons, showing how a uniform optical pattern within a cavity can lead to the formation of solitonic structures. These states were found to be stable but we note that in general the shape of a soliton can evolve, as solitons can be non-stationary. This is the case for Peregrine breather solitons~\cite{Peregrine1983}, exploding solitons~\cite{Knox1992}, which break up then recovering their original shape, chaotic solitons~\cite{SotoCrespo2002}, which remain localized despite randomly changing shape, and bio-solitons~\cite{Rozanov2003}, whose evolution into a labyrinth structure resembles the life cycle of unicellular organisms.

Despite dating back to 1983,  this foundational study includes several ingredients of modern soliton technology, being the utilization of a ring cavity and third-order Kerr nonlinearity. However, while the ring cavity is a common setup for studying cavity solitons, it is not always indispensable. Recent research has shown that standing wave cavities, like the Fabry-Perot, are also capable of supporting solitons and are gaining attention, particularly in the realm of soliton microcombs~\cite{Kippenberg2018DissipativeMicroresonators,Cole2018}.

\subsection{Conservative vs. dissipative solitons}

While cavities are often used to produce solitons, as in the study by Mc Laughlin and co-workers just described, they are not strictly necessary. When solitons propagate in an infinite Kerr medium, such as an optical fiber, without any external perturbations or losses, they are termed conservative solitons. These solitons conserve energy as they travel because they are not subject to external forces or constraints that would add or remove energy from the system.

Conservative solitons are governed by the nonlinear Schrödinger equation (NSE), which can be expressed in its adimensional version as follows \cite{Lugiato_Prati_Brambilla_2015}:

\begin{equation}
i\frac{\partial E}{\partial t} + \frac{\partial^2 E}{\partial x^2} + s_1 |E|^2E = 0
\end{equation}

Here, \( E \) represents the complex envelope of the wave, \( t \) is time describing evolution in the plane transverse to the propagation direction, and \( x \) is the transverse spatial dimension. The equation encapsulates three terms: the first term describes the evolution of the wave along the propagation direction, the second term represents diffraction, and the third term captures the effect of nonlinearity in the medium. Depending on the sign of $s_1$, the NSE allows for two types of solitons: bright ($s_1 = +1$, ``focusing'' case) and dark ($s_1 = -1$, ``defocusing'' case) solitons. Bright solitons are localized increases in intensity against a low background, whereas dark solitons are dips or voids in intensity within a continuous wave background. The NSE has also been used to study hydrodynamic rogue waves---unusually large and unpredictable surface waves observed in the ocean. Remarkably, their optical counterparts appear to have a connection with optical solitons \cite{Solli2007,Chowdury2023}, and the measurement techniques developed for the study of rogue waves are being explored for the characterization of soliton systems \cite{Dudley2019}. 

Introducing a cavity into the system adds complexity. The presence of an injection port in the cavity brings losses, making the system dissipative. Unlike conservative solitons, dissipative solitons require either gain or a continuous input of energy to counteract the losses inherent in the system. Dissipative solitons driven by an external field are described by a modified version of the NSE, known as the damped-driven NSE (here in the ``focusing'' case):

\begin{equation}
\centering
i\frac{\partial E}{\partial t} + \frac{\partial^2 E}{\partial x^2} + |E|^2E = i \varepsilon(- E - i\theta E + E_I)
\end{equation}

In this equation, the right-hand side introduces perturbing terms corresponding to loss (\( -E \)), cavity detuning (\( -i\theta E \)), and the driving field (\( E_I \)), where the magnitude of the perturbation is controlled by $\varepsilon$. In the context of optical pattern formation, this equation (with $\varepsilon=1$) is known as the seminal Lugiato-Lefever equation~\cite{Lugiato1987} (LLE). Unlike the approach used in the study by Mc Laughlin, which required alternating the propagation of the field in the cavity with the coherent addition of the input at each roundtrip, this mean-field model averages the effects over the cavity by considering a uniform field distribution in the propagation direction, removing the need to consider the field propagation explicitly and allowing to use a single partial differential equation with a driving term.

For systems with higher-order nonlinearities, different equations come into play, such as the cubic-quintic complex Ginzburg-Landau equation (CGLE), which has no injection, and the Swift-Hohenberg equation. These models account for more complex interactions and phenomena that can occur in such systems, extending beyond the scope of the standard NSE \cite{ankiewicz2008dissipative,akhmediev1997nonlinear}.

\subsection{Spatial, temporal and spatiotemporal solitons}

Solitons can manifest in various dimensions, with each type presenting unique properties and potential applications \cite{boardman1998temporal}. For instance, the solitons studied by Mc Laughlin are two-dimensional, consisting of spatial optical patterns in the transverse plane. Such spatial solitons have been the subject of extensive study due to their potential for spatial manipulation of information~\cite{Guetlich2007,Cleff2008}. These solitons are localized structures that can act as bits in all-optical memories, capable of being dynamically written and erased using assisting optical pulses~\cite{Barland2002,Firth2002,Hachair2004}. However, the feasibility of using spatial solitons in the visible in practical devices has been challenged by their size. In optical systems, solitons typically span an order of 10 $\mu$m or more, which is considerably larger than bits in conventional hard drives, thereby reducing the possible information density.

Spatial solitons also have the intriguing property of being steerable, allowing them to be repositioned, which was considered attractive for applications such as all-optical delay lines that buffer information packets when a router is overloaded. Another innovative application that was proposed for spatial solitons is soliton force microscopy~\cite{Pedaci2008}, which leverages the sensitivity of solitons to phase perturbations to characterize features such as the curvature of an optical trap.

Over time, there has been a shift within this scientific community from spatial to temporal solitons. Temporal solitons are inherently one-dimensional; they exist in environments like laser cavities or optical fibers, where only the longitudinal dimension is significant. Here, nonlinearity is balanced by dispersion (instead of diffraction as in 2D spatial soliton systems). Temporal solitons have given rise to key technological advancements \cite{dudley2023fifty,Herr2013TemporalMicroresonators}. They have enabled a range of applications, from telecommunications and metrology to medical imaging and photonic computing. These applications exploit the unique properties of temporal solitons, such as their stability and precision in timing, making them suitable for high-precision tasks. Temporal solitons will be our key focus in the following sections of this review.

Finally, spatiotemporal solitons represent the three-dimensional class of solitons, exhibiting localization in both space and time \cite{Columbo_2006,Javaloyes2016,Gustave2017}. In certain configurations, such as nonlinear resonators with saturable absorbers, these are termed ``cavity light bullets''~\cite{Brambilla2004}. The rising field of spatiotemporal optics~\cite{wright2022physics,wright2022nonlinear,shen2023roadmap,yessenov2022space} and highly multimode photonics~\cite{piccardo2021roadmap} is driving new interest in these 3D solitons \cite{Rosanov2001}. These areas leverage correlations between spatial and temporal degrees of freedom of light, leading to the exploration of complex phenomena and potential new applications for spatiotemporal solitons. These solitons could unlock new avenues in high-capacity communications, advanced computing architectures, and more precise measurement techniques, promising to play a central role in the next generation of photonic technologies \cite{Firth2010}.

\section{Classification of soliton sources}







In the intricate landscape of soliton sources, devising a clear-cut classification is a complex task due to the many criteria one could consider. Distinctions can be made based on the nature of modelocking, the type and size of the platform, the emission range, and much more, reflecting the diversity and richness of soliton-generating systems. For the purpose of clarity and coherence in this review, we propose to categorize soliton sources into three broad classes: \textbf{modelocked lasers}, \textbf{passive resonators}, and \textbf{active resonators}. We will restrict our overview only to sources of temporal solitons.

Modelocked lasers and passive resonators represent the more traditional and well-established types of sources, while active resonators are an emerging frontier, characterized by gain (as in a modelocked laser) but also having the possibility of operation under hybrid driving schemes, such as with coupling to a passive cavity or under external injection. In modelocked lasers, the energy that sustains the solitons is entirely internal to the system. This class includes various kinds of lasers, like fiber and solid-state lasers, which can feature different modelocking techniques. On the other hand, passive resonators do not have gain and rely entirely on external energy to create solitons. The distinction between these systems is essential, as it underscores the source of energy that maintains the dissipative soliton within the system.

\subsection{Modelocked lasers}
Modelocked lasers hold a central place in the domain of soliton sources. They were the first system where temporal solitons were not only theoretically predicted but also empirically observed. The mechanism that sustains these solitons is inherently internal, with the laser medium providing essential gain to balance out any intrinsic losses. This internal energy dynamic leads to a steady generation of solitons without the need for external optical driving signal.

Modelocking in lasers can be implemented through a variety of methods~\cite{Gao2020,haus2000mode}, including passive and active techniques. Passive modelocking typically utilizes saturable absorbers, such as semiconductor saturable-absorber mirrors (SESAMs). Active modelocking can involve modulators such as acousto-optic or electro-optic modulators, which are inserted into the cavity to impose a fixed modulation on the laser output.

Historically, the theoretical groundwork for temporal solitons in fibers was laid down by Hasegawa and Tappert in 1973 \cite{Hasegawa1973}, who first predicted their existence. It wasn't until 1980 that Mollenauer and co-workers provided experimental evidence of these solitons, marking a milestone for soliton sources \cite{Mollenauer1980}. 

Modelocked lasers have been utilized in a broad spectrum of configurations, ranging from fiber lasers to Ti:sapphire lasers known for their extensive tuning range. In fiber lasers, the linear birefringence inherent in the fiber's structure caused by stress, imperfections, or bending within the fiber can lead to pulses traveling at different group velocities. However, thanks to a nonlinear index contribution, solitons in orthogonal polarization modes can overcome this and propagate as a cohesive unit, forming what is known as a vector soliton \cite{Christodoulides1988,Menyuk1987}.

\subsection{Passive resonators}

\begin{figure}[!b]
    \centering
    \includegraphics[width=1\textwidth]{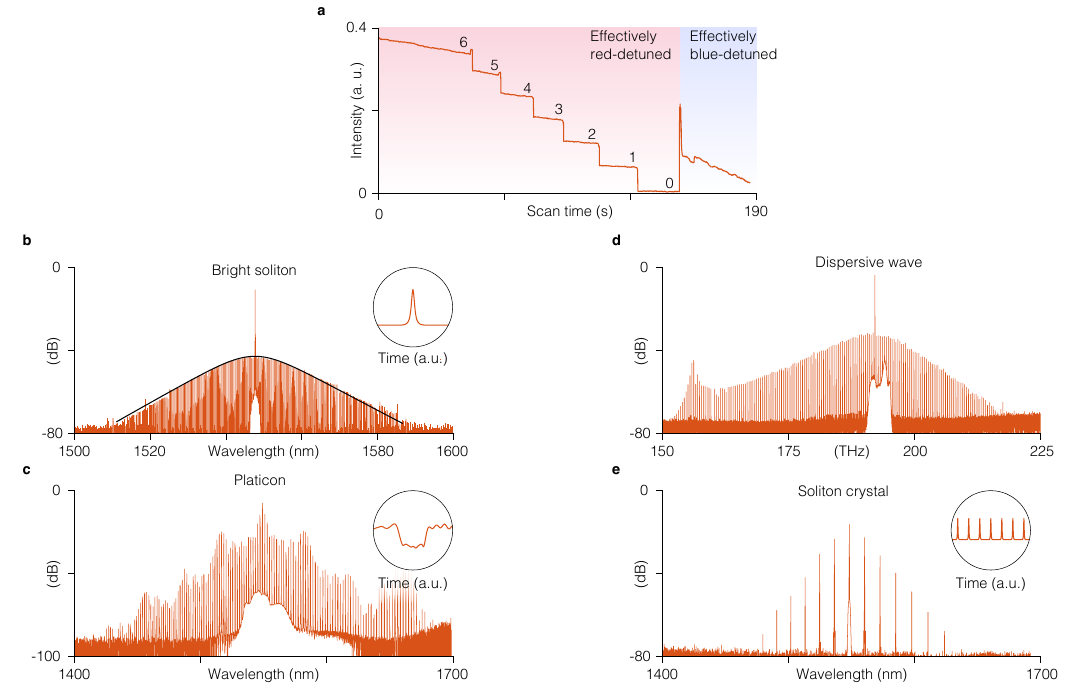}
    \captionsetup{width=1\textwidth}
    \caption{\textbf{a}. Integrated soliton intensity as the wavelength of the punp laser is scanned across the microcavity resonance--a typical soliton generation scheme. A characteristic ladder in the converted average light intensity (``soliton steps") follows the process of the elimination of the solitons (here from six to zero), one by one, as the scan proceeds. \textbf{b-e}. The variety of the achievable soliton states in Kerr microresonators both in the anomalous and the normal dispersion regimes, as shown by their characteristic spectra and waveforms (insets). Data sources: \cite{Kippenberg2018DissipativeMicroresonators,Karpov2019DynamicsMicroresonators,Lobanov2017}.}
    \label{fig:microcombs_solitons}
\end{figure}

\begin{figure*}[!t]
    \includegraphics[clip=true,width=\textwidth]{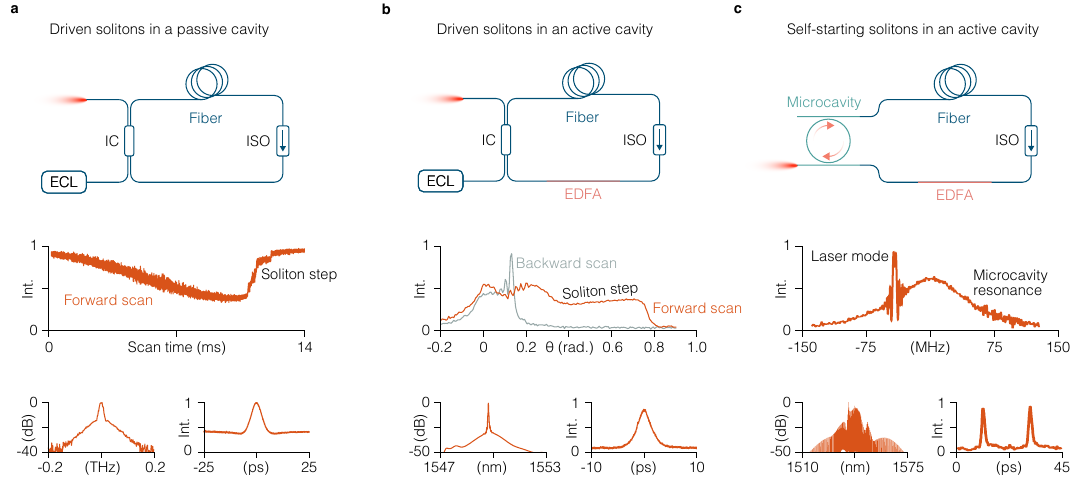}
    \caption{\textbf{Soliton generators based on passive and active fiber resonators.} The columns display the schematics of the three different fiber resonator systems (top row), the soliton excitation scheme (middle row), and the spectrum and the temporal profile of the resulting soliton state. ECL, external cavity laser. IC, input coupler. ISO, optical isolator. EDFA, erbium-doped fiber amplifier. \textbf{a}, Dissipative cavity solitons, first obtained in driven passive fiber cavities, emerge from the instability of an externally coupled CW wave. One possible excitation scheme involves scanning the wavelength of a tunable laser from the blue-detuned side to the red-detuned side of the cavity resonance. The appearance of the soliton is signified by a characteristic step in the intensity of the transmitted laser beam. The soliton spectrum features a strong pump mode in its center --- the background driving laser field, also visible in the temporal profile of the waveform circulating inside of the fiber resonator. \textbf{b}, Here, the fiber loop incorporates an optical amplifier (EDFA), yet the cavity remains below its lasing threshold. The purpose of the amplifier is to raise the cavity quality factor, relaxing the power requirements on the drive laser for soliton generation. Data adapted from ref.~\cite{Englebert2021TemporalResonator}. \textbf{c}, Unlike the systems in \textbf{a} and \textbf{b}, in this one there is no external pump laser. The solitons emerge spontaneously by driving the cavity above the lasing threshold. An additional microcavity resonator, embedded as a part of the fiber cavity, provides the necessary nonlinearity for soliton generation. The lasing mode appears red-detuned with respect to the microcavity resonance --- the condidition necessary for the soliton stability. Data adapted from ref.~\cite{Rowley2022Self-emergenceMicrocavity}.}
     \label{fig:active_passive_fiber}
\end{figure*}

Passive resonators, implemented in fiber optic systems (Fig.~\ref{fig:active_passive_fiber}\textbf{a}) and in particular their microscale counterparts \cite{DelHaye2007OpticalMicroresonator,Kippenberg2018DissipativeMicroresonators}, have become a cornerstone in the study and application of soliton physics. These devices harness the power of external continuous wave pumps to incite soliton formation within the cavity through a balance between their intrinsic nonlinearity and dispersion.

Microresonator-based frequency combs---often referred to as microcombs---are based on parametric gain and operate through a process called Kerr frequency comb generation. The tuning into the soliton regime in microresonators is a nuanced process \cite{Herr2012UniversalMicroresonators}. It involves carefully adjusting the pump laser wavelength to achieve bistability within the system, a prerequisite for soliton formation. The continuous wave pump light, when tuned into resonance with the microresonator modes, undergoes nonlinear four-wave-mixing, leading to the generation of new optical frequencies. The interplay between this parametric gain, dispersion, and nonlinearity in the cavity results in the formation of solitons~\cite{Kippenberg2018DissipativeMicroresonators}.

The potential applications of microcombs range across various fields. They are pivotal in the calibration of astronomical spectrometers, where their natively large intermode spacing in the range of tens to hundreds of gigahertz can help in measuring the redshift of distant cosmic objects. In the realm of telecommunications, microcombs may enable massively parallel coherent communications, enhancing data transmission rates~\cite{Marin-Palomo2017Microresonator-basedCommunications}. Their use in ultrafast distance measurements is revolutionizing LIDAR and related technologies, providing higher resolution and accuracy. The high stability and large optical bandwidth of microcombs also makes them attractive for optical atomic clocks. Another cutting-edge application is dual-comb spectroscopy, which uses two microcombs for high-precision molecular spectroscopy, allowing for rapid environmental sensing and analysis.

Passive microresonators are not just devices for practical applications but are also a rich playground for the experimental studies of the complex soliton physics (Fig. \ref{fig:microcombs_solitons}). They can support a variety of soliton states \cite{Herr2013TemporalMicroresonators}, such as bright solitons that correspond to localized peaks of light, and dark solitons, which are intensity dips in a continuous wave background. Microresonators can also give rise to exotic states like Cherenkov dispersive waves and Raman solitons, where Raman scattering effects contribute to soliton formation and dynamics. Soliton crystals \cite{Karpov2019DynamicsMicroresonators}, where solitons are arranged in a fixed periodic structure, and platicons \cite{Lobanov2017}, flat-top solitons, are among the complex solitonic structures observed in these systems.

Recent advancements have emphasized the shift from conventional ring geometries to Fabry-Perot cavities with Bragg reflectors \cite{Wildi2023,Cole2018,Yu2019}, presenting the advantage of decoupling the device layer configuration from dispersion engineering. Efforts are also being concentrated on integrating laser sources with passive microresonators on chips \cite{Stern2018,Shu2022Microcomb-drivenSystems}. This miniaturization and integration are pushing the boundaries of what's possible with soliton physics and applications, paving the way for more compact, efficient, and versatile devices. With their ability to host a multitude of solitonic phenomena and their growing list of real-world applications, passive microresonators are set to remain at the forefront of research and development in photonics.

\subsection{Active resonators}
Active resonators represent an expanding frontier in soliton research. These resonators possess gain, allowing them to actively sustain solitons, but can also be used with hybrid driving schemes, such as with coupling to a passive cavity or under external optical injection---while being driven either below or above the cavity lasing threshold. This differentiates this class of soliton sources from conventional modelocked lasers and endows it with additional control and versatility.

Active fiber resonators have shown novel solitonic properties, where coherent driving and incoherent pumping merge to produce high-power, ultra-stable pulse trains (Fig.~\ref{fig:active_passive_fiber}\textbf{b}) \cite{Englebert2021TemporalResonator}. Such systems exhibit high-peak-power solitons on a low-power background and present amplified spontaneous emission that has a negligible impact on soliton stability, with advantages for metrology, spectroscopy, and communications.

Microresonator-filtered fiber lasers are another exemplary hybrid system \cite{Rowley2022Self-emergenceMicrocavity,Bao2019LaserMicrocombs,Nie2022DissipativeLasers}. They harness slow nonlinearities of the system, transforming temporal cavity solitons into the system's dominant attractor. Such solitons self-start and can recover spontaneously even after being disrupted, offering reliable and robust soliton microcombs (Fig.~\ref{fig:active_passive_fiber}\textbf{c}). These systems can exhibit broad bandwidths and high mode efficiency, and the study of their real-time soliton formation and interaction dynamics has been facilitated by ultrahigh-speed time magnification techniques.

There has been a notable development in turnkey operation for soliton microcombs co-integrated with a pump laser \cite{shen2020integrated}. These microcombs can immediately generate solitons upon activation of the pump laser, bypassing the need for complex startup and feedback protocols. The key physical mechanism enabling turnkey operation in this system is the presence of feedback from the microresonator into the pump laser, which results in spontaneous self-tuning into the soliton regime.

Lastly, driven semiconductor ring lasers, which we will explore in greater detail later in this review, have successfully produced phase solitons in macroscopic cavities \cite{Gustave2015} and bright solitons in microscopic cavities~\cite{kazakov2024driven,Columbo2021UnifyingLasers}. The latter were demonstrated in the mid-infrared wavelength range, a region that has eluded compact soliton sources until now. These continuous-wave-driven semiconductor laser chips rely on a fast bistability akin to that found in passive nonlinear Kerr resonators, offering a promising route to robust and industrially manufacturable nonlinear integrated photonics in the mid-infrared.

\medskip

These diverse examples illustrate the multifaceted nature of active resonators with hybrid driving schemes. They combine traditional active laser dynamics with innovative control mechanisms, paving the way for the next generation of soliton sources with broadened capabilities and applications.

\section{Solitons in quantum cascade lasers}

\subsection{Key properties of QCLs}

In what follows we will consider, almost exclusively, a new emerging platform for soliton generation in active resonators based on one particular semiconductor laser gain region --- a quantum cascade laser (QCL). First demonstrated experimentally in 1994 at AT\&T Bell Laboratories~\cite{faist1994quantum}, QCLs have since become the dominant source of coherent radiation in the mid- to long-wave infrared part of the electromagnetic spectrum ($3-30$ $\upmu$m). QCLs are renowned for their ultrafast (subpicosecond) gain recovery time, a feature that recently turned out to be central to their ability to generate optical solitons. Unlike conventional semiconductor lasers that utilize interband transitions (and the gain recovery time of several nanoseconds), QCLs rely on intersubband transitions within the conduction band of semiconductor materials. These intersubband transitions inherently possess faster relaxation times as they involve transitions between subbands within the same conduction band due to fast optical phonon scattering, rather than interband ones which occur across conduction and valence band. Intersubband and intrasubband relaxation times range from sub-picoseconds to picoseconds; interband transitions instead are characterized by much longer relaxation due to carrier recombination. The architecture of QCLs, comprising a series of coupled quantum wells and barriers forming a superlattice, facilitates rapid electron recycling. After emitting a photon and transitioning to a lower energy subband, electrons quickly tunnel through the barrier into the upper laser state of the next stage, thus rapidly availing themselves for subsequent photon emission.

As the laser reaches and surpasses threshold, the contribution to gain recovery stemming from the finite lifetime of the upper lasing state within the active region greatly diminishes owing to the onset of stimulated emission. Thus, electron drift is controlled by the density of cavity photons. Gain recovery occurs as a result of the transport of electrons across the active regions and the interconnecting regions and occurs on a time scale of in the 1-10 ps range \cite{Choi2008}. The nature of gain recovery deviates from that of traditional lasers due to the inclusion of superlattice transport in the cascade.

It has been shown theoretically that the fast nonlinearity of QCLs~\cite{Piccardo2018Time-dependentCombs,Piccardo2019LightPhenomena} that couples the fluctuations of the amplitude of the intracavity field to the fluctuations of its phase, arises from the Bloch gain~\cite{Opacak2021FrequencyNonlinearity,Wacker2007}. The phase-amplitude coupling does not vanish even at high modulation speeds (tens to hundreds of gigahertz). At low modulation speeds (kilohertz) this amplitude to phase coupling is not unique to QCLs, but is a property shared by all semiconductor lasers, and is widely known as the linewidth enhancement, quantified by the linewidth enhancement factor (LEF) $\alpha$ (also known as Henry factor~\cite{Prati2007}). In most semiconductor lasers it is an undesired, yet unavoidable effect that leads to the broadening of the laser linewdith by $(1+\alpha^2)$ beyond the Shawlow-Townes limit. However, in QCLs, the fast optical response of the gain medium --- several orders of magnitude faster than the typical roundtrip time in the laser cavity ---  leads to what can be seen as an effective Kerr nonlinearity --- an intensity-dependent refractive index --- quantified by the LEF, very much in the spirit of the Kerr nonlinearity in passive microresonators and optical fibers. Furthermore, the LEF in QCLs can be modified by engineering the optical gain~\cite{terazzi2007bloch}, is a dynamic property that itself depends on the intracavity field intensity~\cite{pilat2023hot}, and is naturally dispersive (LEF is strongly wavelength-dependent across the laser operating bandwidth)~\cite{Opacak2021SpectrallyComb}. In the case of QCLs, this nonlinearity is resonant with the optical transitions in the gain region, therefore the associated third-order susceptibility $\chi^{(3)}$ is several orders of magnitude larger than the $\chi^{(3)}$ of silica or Si$_3$N$_4$, of which fibers and microresonators are typically made. The giant resonant $\chi^{(3)}$ gives rise to the strong four-wave mixing effects in a QCL above threshold~\cite{Friedli2013Four-waveAmplifier}, and is behind the formation of frequency-modulated (FM) combs in Fabry-Perot QCLs~\cite{Hugi2012Mid-infraredLaser}. More recently, the discovery of the link between the LEF and the Kerr nonlinear index, and the associated dynamic effects, triggered a new wave of research in instabilities and frequency comb formation in unidirectional ring cavity QCLs~\cite{heckelmann2023quantum,Columbo:18}, and of soliton generation in these systems. It is the nonlinear dynamics of the laser states in ring cavity QCLs that will be the focus of the remaining sections of this review.   

\subsection{A change in cavity geometry enables soliton generation}

The route from single-mode operation of the laser just above its first threshold to the second threshold (also known as an instability threshold) where the laser output becomes multimode is paved by the mechanisms that render the single-mode solution unstable. First found in Fabry-Perot QCLs~\cite{Mansuripur2016Single-modeOscillator}, the single-mode instability is caused by the combination of the unclamped Lorentzian gain that arises due to the populating grating in a standing-wave bidirectional cavity --- a phenomenon known as spatial hole burning (SHB) --- and a parametric gain due to population pulsations (PPs). SHB leads to the increase of the gain with increasing pumping for the non-lasing longitudinal modes above the first threshold, favoring the modes nearest to the primary one. PPs provide a parametric contribution to the gain of the more distant sidebands by coherent temporal carrier oscillations at the frequencies of the beat notes between the primary lasing mode and the sidebands. The combination of the SHB and PPs leads to the onset of dense frequency combs with the intermode spacing of one cavity free spectral range (FSR)~\cite{faist2016quantum, Piccardo2022LaserApplications} and harmonic frequency combs with the intermode spacing of several FSRs~\cite{Mansuripur2016Single-modeOscillator, Kazakov2017Self-startingLaser,Piccardo2018TheApplications}. 

The standing-wave nature of the Fabry-Perot cavity has long been thought of as essential to the observation of single-mode instability and generation of phase-coherent frequency combs states in QCLs at experimentally achievable pumping levels. Removing the contribution of the SHB to the total sideband gain has long been postulated to lead to a dramatic increase in the instability threshold, requiring the pumping levels at least nine times above the first laser threshold~\cite{Mansuripur2016Single-modeOscillator} --- pumping level, rarely achievable in state-of-the-art ridge Fabry-Perot QCLs. Contrary to this notion was the experimental observation of frequency comb states generated by unidirectional ring QCLs, where SHB is suppressed, almost simultaneously by two research groups~\cite{Meng2020Mid-infraredLaser,Piccardo2020FrequencyTurbulence}. The frequency comb states have been observed at the pumping levels only fractionally higher than the first laser threshold. While there is an ongoing debate on whether the remaining SHB due to a weak back-reflection of the unidirectional field in a ring cavity may play a role in lowering the instability threshold~\cite{seitner2024backscattering}, the LEF has been shown to play a key role in the onset of the instability~\cite{opavcak2024nozaki,Columbo:18}. A self-consistent theory based on coupled Maxwell-Bloch equations, where the LEF is accounted for, can predict the onset of the instability at moderate pumping levels, and furthermore, was shown to be reducible to a complex Ginzburg-Landau equation (CGLE) --- a model vastly studied in the areas of hydrodynamics and spatially correlated quantum systems (Bose-Enstein condensates). Among the solutions to the CGLE is the special kind of soliton --- the Nozaki-Bekki soliton, that has been recently observed experimentally in ring QCLs~\cite{opavcak2024nozaki}.

\begin{figure}[!h]
    \includegraphics[width=1\textwidth]{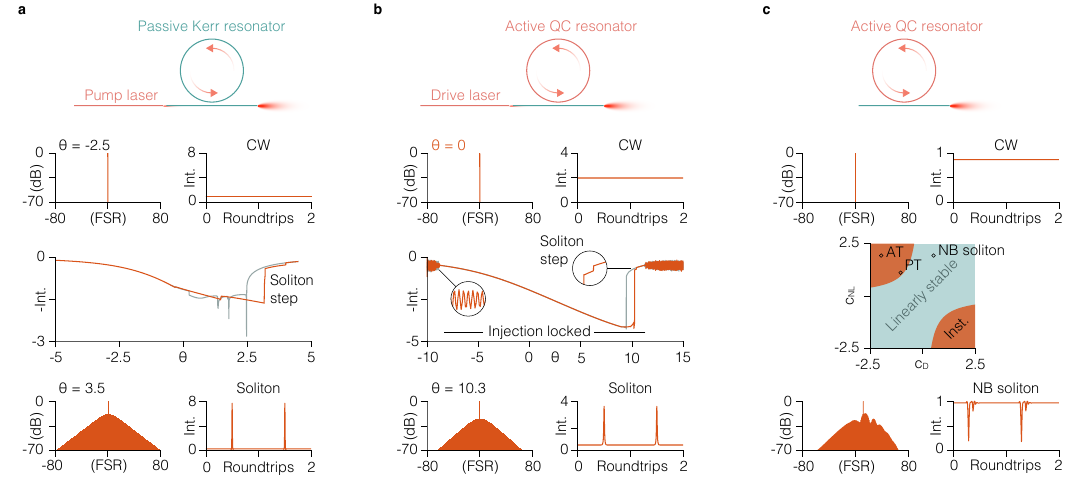}
    \captionsetup{width=1\textwidth}
    \caption{\textbf{a}. Passive Kerr resonator optically pumped by an external laser experiences a bistable behavior in the output intensity as a function of the detuning $\uptheta$ of the laser wavelength from the Kerr cavity resonance. Here the middle panel shows the mean intracavity intensity as function of detuning. The intensity values are artificially flipped by multiplying with $-1$ to facilitate the visual comparison with the experimental plots later in the manuscript, where not the intracavity intensity, but the measured output intensity is shown. On a forward scan (blue-detuned to red-detuned, orange curve in the middle panel) the resonator transmission shows an abrupt step signifying generation of one or several cavity solitons, which cannot be generated on a backward scan (gray curve in the middle panel). The simulations are based on the GLLE. \textbf{b}. Active quantum cascade (QC) resonator with an on-chip integrated drive laser. The soliton excitation scheme is similar to that of passive Kerr resonators and entails scanning the frequency of the drive laser through the lasing resonance of the active racetrack resonator. The associated output intensity (orange curve --- forward scan, gray curve --- backward scan) exhibits optical bistability and a characteristic step in transmission when generating a soliton on a forward wavelength scan. \textbf{c}. Free-running active QC resonator in the absence of an external drive laser can be described by the complex Ginzburg-Landau equation (CGLE). The instability of the primary lasing mode caused by phase turbulence (PT) or amplitude turbulence (AT) may lead to the generation of homoclons. Additionally, dark solitons solutions (Nozaki-Bekki optical solitons) can emerge in the linearly stable region of the CGLE. The figure is partially adapted from Refs.~\cite{Columbo2021UnifyingLasers,kazakov2024driven}.} 
     \label{fig:active_passive_theory}
\end{figure}

\subsection{From the Ginzburg-Landau to the generalized Lugiato-Lefever equation}
Unidirectional field propagation inside of the ring QCL cavity (either clockwise or counterclockwise) and the fast gain recovery time allow a drastic simplification of the complete Maxwell-Bloch equations (MBE) laser model. This is achieved by eliminating half of the equations for the backward propagating wave amplitude, and by making the approximation of adiabatic carrier dynamics. The complete derivation of the CGLE from the full MBE model is given in Ref.~\cite{Piccardo2020FrequencyTurbulence,Columbo2021UnifyingLasers}.
The CGLE shows that the instability experimentally measured in ring QCLs originates from phase turbulence, leading to the formation of localized structures thanks to the interplay of dispersive and nonlinear effects, as in the modulational instability of passive microresonators.

Following these experimental observations there was an effort to unify previously disparate theoretical descriptions of comb generation in optically driven passive nonlinear resonators and active ring resonators, with or without an external optical pump. This led to the formulation of a generalized Lugiato-Lefever equation (GLLE, Fig.~\ref{fig:active_passive_theory}\textbf{a})~\cite{Columbo2021UnifyingLasers} 
\begin{equation}
\label{eq:gen}
    \tau_p\:\partial_t E = E_I + \left(-1-i\theta_0\right)E + \left(d_R+id_I\right)\partial_z^2E\\ + \mu\left(1-i\Delta\right)\left(1-|E|^2\right)E,
\end{equation}
that describes the spatiotemporal evolution of the electric field envelope in an optical cavity, that can be either filled with a nonlinear lossy medium or contain optical gain. The sign of the gain/absorption parameter $\mu$ in the GLLE controls the nature of the medium --- passive or active. Variables in GLLE are as follows: $E_I$ --- driving field; $\tau_p$ --- photon lifetime; $\theta_0$ --- normalized pump detuning; $\Delta$ --- third order nonlinear coefficient; $\mu$ --- gain/absorption parameter; $d_R$ and $d_I$ --- diffusion and dispersion parameters. The GLLE allows to transfer the established methods of waveform control from passive Kerr microresonators into the realm of active resonators. In Kerr resonators by an adjustment of the optical pump intensity and its detuning from the cavity resonance it is possible to control the waveform generated inside the resonator. In what follows we show how the GLLE predicts that --- in a similar fashion --- it is possible to switch between the different states of light emitted by QCL ring resonators, bringing a new level of control over the waveforms emitted by these lasers. We consider three regimes of the GLLE describing the behavior of three different resonator systems.

\textbf{Passive resonators with an external optical drive}. With the medium being weakly absorbing ($\mu<0$, $|\mu|\ll1$) the GLLE reduces to the original LLE that fully captures the dynamics of passive resonator Kerr combs. It predicts the formation of cavity solitons (CSs) and Kerr soliton crystals, as a function of detuning and injected power (Fig.~\ref{fig:active_passive_theory}\textbf{b}). 

\textbf{Active resonators without external optical drive}. In a ring QCL cavity above lasing threshold and in absence of an injected field ($\mu>0$, $E_I = 0$) the GLLE reduces to the complex Ginzburg-Landau equation (CGLE) whose solution space is spanned by the two laser parameters — group velocity dispersion (GVD) and linewidth enhancement factor (LEF). With the appropriate values of GVD and LEF the laser operates in the instability region where the single-mode optical field, destabilized by phase turbulence, will evolve into one of many possible steady-state structured waveforms, such as Nozaki-Bekki solitons, phase solitons, Turing rolls, and homoclons (Fig.~\ref{fig:active_passive_theory}\textbf{d}). 

\textbf{Active resonators above thresholds with an external coherent drive}. Injecting an external control optical field from a wavelength-tunable laser enables a new kind of dynamics in presence of the optical bistability. In such an externally driven scheme, the intracavity field will injection lock to the drive field when the frequency detuning between the two is sufficiently small (Fig.~\ref{fig:active_passive_theory}\textbf{c}). The intracavity field frequency will follow the frequency of the drive field --- as long as the drive field remains within the finite locking range of frequencies --- thus effectively decoupling it from the cavity resonance. Previously it was theoretically proposed that such a scheme would allow the generation of optical solitons in a laser once the intracavity field is injection locked to the drive field and the intensity of the driving field is first increased, and then decreased~\cite{Columbo2021UnifyingLasers}. Alternatively, akin to the soliton generation scheme of Kerr microresonators, the solitons can also be generated by performing wavelength scans of the drive field through the lasing resonance (Fig.~\ref{fig:active_passive_theory}\textbf{c}). The intracavity field injection locks to the drive and generates one or several cavity solitons --- in direct analogy with passive Kerr microresonators.


\begin{figure*}[!t]
    \includegraphics[clip=true,width=\textwidth]{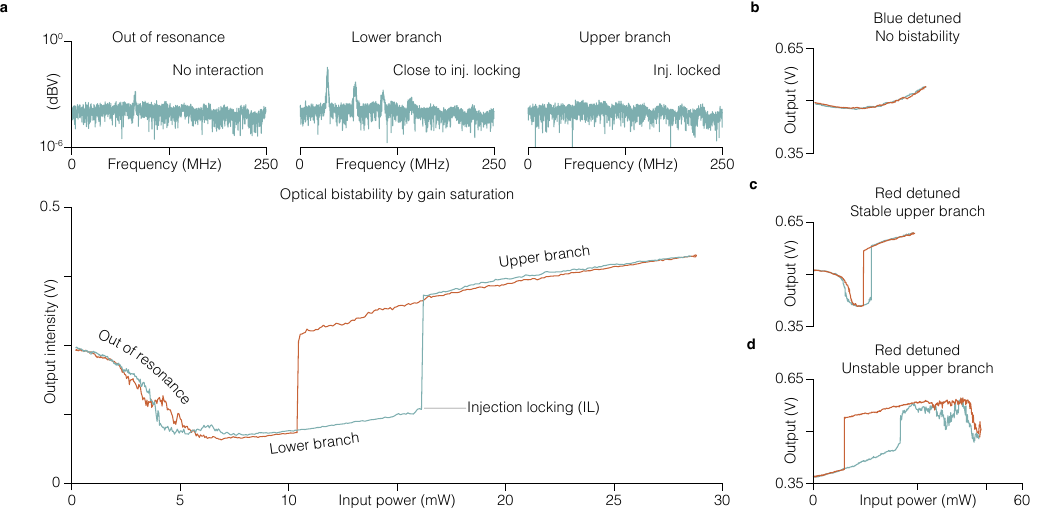}
    \caption{Optical bistability in a driven racetrack QCL induced by varying the incident optical power. \textbf{a}. Output power of the driven RT resonator above the threshold as a function of the drive laser input power, when it is 150 MHz red detuned from the RT frequency. Bistable behaviour emerges when first increasing the drive power (cyan), and then decreasing it (red). Top panels show the spectra of the heterodyne beat note between the RT and the drive laser. \textbf{b}. Output power for 200 MHz blue detuning. \textbf{c}. Output power in case of 200 MHz red detuning. \textbf{d}. Output power in case of 200 MHz red detuning for a higher bias setpoint of the RT. Figure taken from~\cite{kazakov2024driven}}.
     \label{fig:SI_bistability_power}
\end{figure*}

\subsection{Design of ring QCLs with an optical injection port}

Efficient injection of an external optical drive signal into a QC ring resonator, as well as the efficient extraction of internally generated light, requires special consideration for the coupler design. An evanescent wave directional coupler is the natural choice in that it does not induce additional back-scattering inside of the ring resonator above threshold, maintaining the unidirectional lasing --- the prerequisite for the soliton generation, as postulated in the GLLE theory. Ring resonators were constructed in the shape of racetracks (RT) on an indium phosphide (InP) substrate, which was epitaxially layered with active material \cite{kazakov2024active}. These were developed using a conventional ridge technique and featured integrated evanescent wave directional couplers, where the coupler waveguide (WG) is narrowly spaced from the RT's straight section by an air gap. The design stage determines the power coupling coefficient through the interaction region's length and the air gap's width. Specifically, with a gap width of $0.8$ \si{\micro\meter} achievable by optical lithography, a simulated power coupling ratio of -10 dB was obtained. The produced devices displayed gaps that were slightly underetched, resulting in a coupling coefficient that exceeded the intended design value. Both the WG and RT incorporate the QC gain medium and are equipped with distinct contacts for separate electrical activation. This configuration of the WG coupler being active provides dual advantages. First, it enables precise adjustment of the coupling intensity through independent current modulation to the WG and RT, which causes a discrepancy in mode indices between them. Additionally, when the WG is powered beyond its transparency current, it functions as a standalone optical amplifier, enhancing the input and output waves by either offsetting insertion losses or delivering overall amplification.


\subsection{Optical bistability in coherently driven ring QCLs}

The bistability can be observed in externally driven ring QCLs while fixing the detuning of the drive laser from the ring laser frequency and sweeping the power of the drive laser. Here the input power of the drive laser is such that the RT is in the regime of strong gain saturation. Due to the non-zero linewidth enhancement factor, saturation of the gain --- which is proportional to the imaginary part of the optical susceptibility $\Im\{\chi\}$ of the active medium --- leads to a change of its real part $\Re\{\chi\}$. Provided the picosecond gain recovery time of the QC active region, this saturation-induced change of $\Re\{\chi\}$ results in an effective fast Kerr nonlinearity, which is slower than the nearly instantaneous passive Kerr nonlinearity, but still leads to an optical bistability in the input-output power relation of the driven RT resonator when the coherent drive laser is red-detuned from the RT's lasing resonance (Fig.~\ref{fig:SI_bistability_power}\textbf{a}), but not when it is blue-detuned (Fig.~\ref{fig:SI_bistability_power}\textbf{b}). Depending on the bias of the RT, the upper branch of the bistability curve can be stable or unstable (Fig.~\ref{fig:SI_bistability_power}\textbf{c}, \textbf{d}).

\begin{figure*}[!t]
    \includegraphics[clip=true,width=\textwidth]{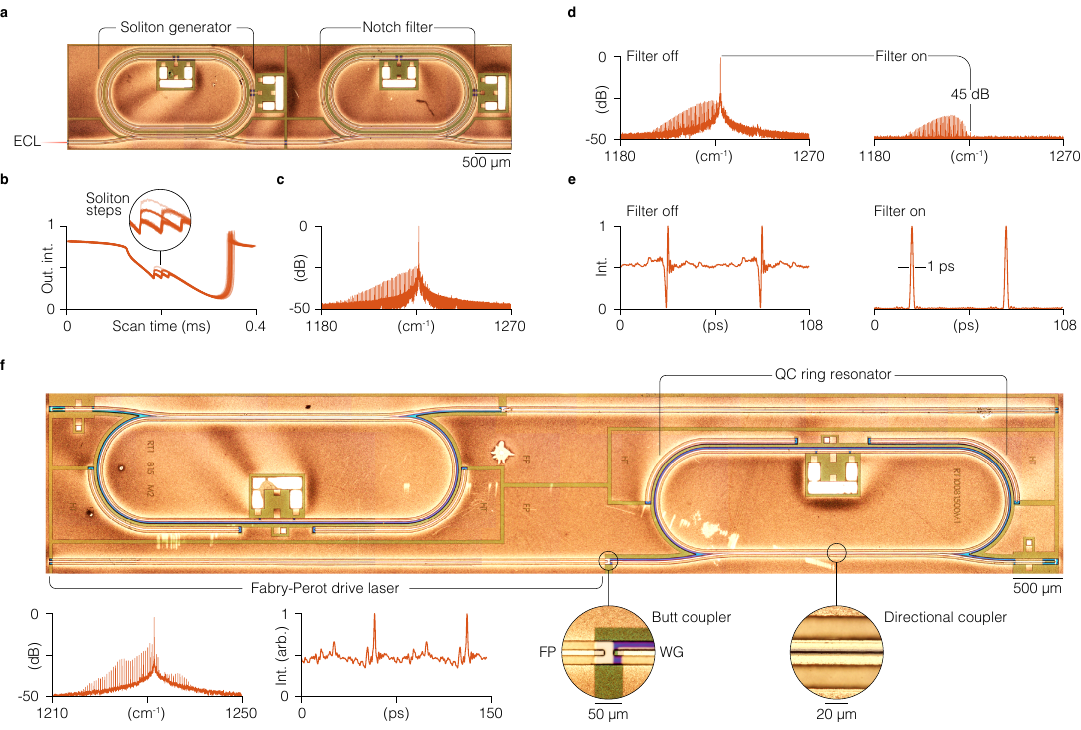}
    \caption{\textbf{Driven solitons in active QC resonators}. \textbf{a}. Optical microscope image of the active QC resonator photonic integrated chip. The chip contains two identical racetrack resonators connected via a bus waveguide --- one for soliton generation, another one --- for drive laser suppression and is driven by an off-chip external cavity laser (ECL). \textbf{b}. Overlaid output intensity curves over one thousand backward scans, showing multistability of the soliton states. The traces are aligned so that the first steps in the output intensity coincide. \textbf{c}. Optical spectrum of the driven soliton. \textbf{d}. Optical spectra of the soliton generator when the filter is switched off and switched on. The suppression of the pump frequency by 45 dB is achieved by fine-tuning the current of the filter and of the integrated heater. \textbf{e}. Reconstructed temporal waveforms of the soliton generator over two consecutive cavity roundtrips with filter off and filter on. \textbf{f}. Optical microscope image of the QC photonic chip with a monolitically integrated drive laser. The chip contains two identical devices each comprised by four components: a Fabry-Perot drive laser (FP), a waveguide coupler (WG), a resistive heater (HT) and a racetrack resonator (RT). Insets show blown up micrographs of the coupling regions between the FP and the WG (butt coupler) and between the WG and the RT (directional coupler). The figure is adapted from ref.~\cite{kazakov2024driven}.}
     \label{fig:integrated_soliton_generators}
\end{figure*}

\section{Outlook on challenges and applications}


\subsection{Turnkey integrated soliton generators}
The use of an external laser for soliton generation and the need for precise optical alignment pose significant challenges for the practical application of these devices. To overcome these issues, a mid-infrared photonic chip can be designed in such a way as to incorporate the pump laser directly with the active racetrack resonator (Fig.~\ref{fig:integrated_soliton_generators}). This integration is exemplified by a design where the drive signal originates from a Fabry-Perot (FP) cavity laser, which is directly coupled to the waveguide directional coupler. By activating both the FP and the racetrack resonator above their lasing thresholds in a single-mode operation and adjusting the current of a nearby heater, it is possible to sweep the wavelength of the FP across the racetrack resonance. During a sweep from a red-detuned to blue-detuned setting, the output intensity shows low-frequency fluctuations, similar to those seen in the modulation instability comb regime of passive Kerr resonators. This regime is characterized by the appearance of multiple frequency tones in the associated RF spectrum of the intermode beat note. Conversely, during a sweep from blue-detuned to red-detuned, the formation of solitons is marked by the cessation of intensity fluctuations and the appearance of a stable microwave tone, indicating a highly phase-coherent multimode state. The soliton state displays a spectrum with tens of mutually locked lines, leading to the emission of a stable bright pulse against a continuous wave (CW) background.

Another key advantage of this integrated soliton generator is its compactness, alongside the long-term stability of the soliton state and its capability to be reliably restored after power cycling. Soliton generation and perfect state recovery through several power cycles is possible by sequentially activating and deactivating the drive currents of the integrated components. Once initiated, the soliton state remains stable over extended periods, demonstrating the robustness and reliability of the fully integrated system. These characteristics are not only significant for the QCL active resonator platform but also reflect qualities found in Si$_3$N$_4$ Kerr resonators that are heterogeneously integrated with III-V semiconductor pump lasers.


\subsection{Perspective of agile waveforms and active control}

\begin{figure*}[!t]
    \includegraphics[clip=true,width=\textwidth]{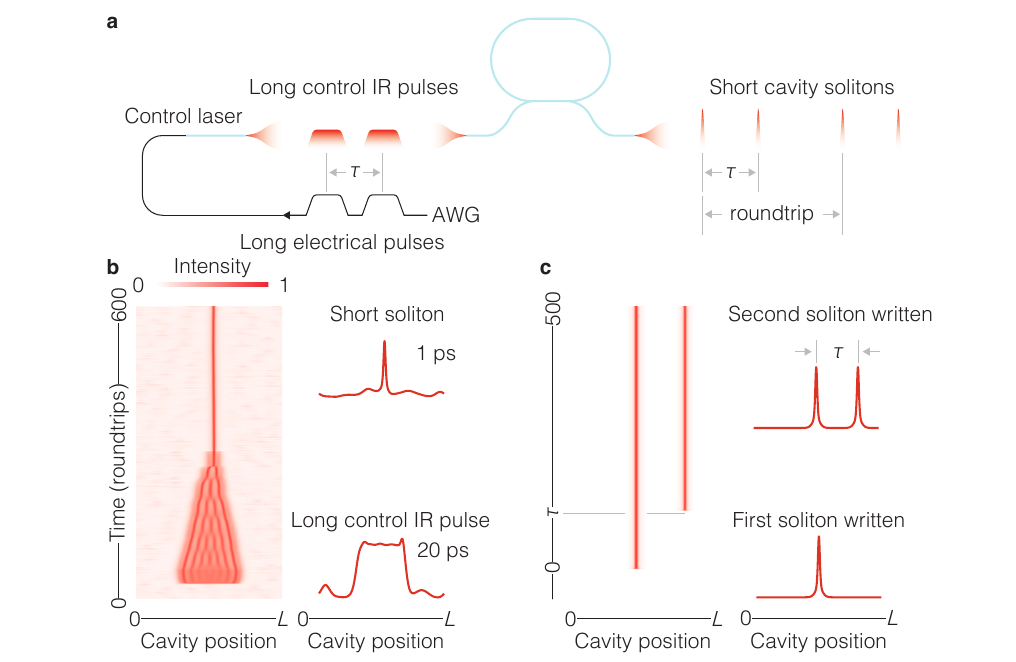}
    \caption{\textbf{Soliton addressing in coherently driven ring QCLs.} \textbf{a}. Arbitrary waveform generator (AWG) injects long electrical pulses with a temporal separation of \(\tau\) into the control laser, modulating its output waveform. The resulting long control IR pulses with the same separation of \(\tau\) are injected into the racetrack QCL triggering generation of short solitons with a separation defined by the time of arrival of the long control pulses. \textbf{b}. Simulated spatiotemporal evolution of intracavity intensity after injection of a long control IR pulse. \textbf{c}. Simulated spatiotemporal evolution of intracavity intensity as two solitons are subsequently written via injection port with a temporal separation of \(\tau\). Simulation results are taken from~\cite{Columbo2021UnifyingLasers}
}
     \label{fig:soliton_addressing}
\end{figure*}

Numerical simulations of the GLLE demonstrate that it is possible to excite solitons by means of long control optical pulses and in doing so to reconfigure the waveform dynamically~\cite{Columbo2021UnifyingLasers}. In this driving scheme long control pulses injected into racetrack QCL get transformed into short solitons as they circulate inside QCL cavity (Fig.~\ref{fig:soliton_addressing}\textbf{a}, \textbf{b}). Long control IR pulses can be generated by direct electrical modulation of the control laser. The numerical simulation in Figure 8b shows how a broad control IR pulse injected into QCL cavity evolves into a short soliton. Arbitrary pulse sequences can be inscribed in a similar manner by injecting a series of control pulses with pulse separation programmed electronically. Fig.~\ref{fig:soliton_addressing}\textbf{c} shows the simulated spatiotemporal field evolution inside the QCL cavity as two pulses are subsequently injected through the input port. While yet to be demonstrated experimentally in active resonators, soliton addressing is a well-known concept in passive resonators~\cite{Leo2010TemporalBuffer}. In case of QCL active resonators such driving scheme would be particularly interesting for the generation of precisely timed pulse sequences with such applications as pump-probe spectroscopy and two-dimensional nonlinear mid-infrared spectroscopy (Fig. \ref{fig:soliton_applications})~\cite{Middleton2010Residue-specificSpectroscopy,Cundiff2013OpticalSpectroscopy}. 

\subsection{Emitting background-free bright pulses with on-chip active ring filters}

A significant challenge with driven cavity solitons is their association with a strong background driving field, which can be $20-30$ dB more intense than the rest of the soliton spectrum. This characteristic necessitates the need for detectors capable of handling a high dynamic range and might lead to the gain depletion in optical amplifiers due to the overpowering laser line.

In contrast to active resonators, passive Kerr resonators often employ methods like external filtering through fiber Bragg grating or the redirection of the intracavity field using an add-drop waveguide to exclude the pump. However, these techniques face limitations; the former is incompatible with photonic integration due to reliance on free-space components in mid-infrared optics, while the latter is ineffective for active resonators where the pump field, essential for soliton formation, is internally generated and would also be emitted through the drop port.

An alternative approach circumvents these issues through the implementation of a photonic integrated notch filter. By using an active ring resonator, operating below the lasing threshold, it is possible to create a filter where the coupling condition, quality factor, and resonance frequency are adjustable independently through electrical pumping (Fig.~\ref{fig:integrated_soliton_generators}\textbf{a}). Precise targeting of a single laser line is possible with a notch filter integrated on the same chip with the soliton generator. Both the soliton generator and the notch filter can be made from identical racetrack resonators connected by a bus waveguide, with finely tunable free spectral ranges through adjustments in pump current and integrated heaters. The soliton is initiated by an off-chip external cavity laser in the first resonator operating above threshold, with the pump line then being filtered by the second resonator, which is operated below threshold. This setup achieves a selective suppression of the pump line by up to 45 dB when aligned with the driving field, demonstrating a highly selective influence on the driving field without significantly altering the rest of the comb spectrum (Fig.~\ref{fig:integrated_soliton_generators}\textbf{d}). The result is a transformation of a soliton from a high-intensity background state to a bright pulse with no background when the filter is activated (Fig.~\ref{fig:integrated_soliton_generators}\textbf{e}).

\subsection{Applications of QCL mid-IR active resonators}

\begin{figure*}[!t]
    \includegraphics[clip=true,width=\textwidth]{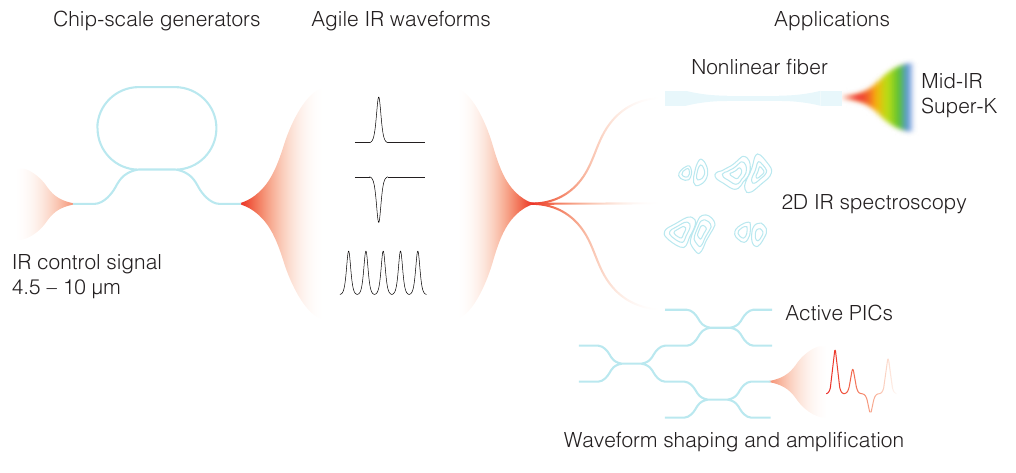}
    \caption{\textbf{Applications of active resonator soliton generators.} Compact chip-scale electrically driven QCLs will generate agile waveforms in the mid-infrared. Applications include nonlinear photonics --- supercontinuum generation and probing of dynamics of fundamental material interactions in the mid-IR range and active photonic integrated circuits (PICs) for arbitrary waveform shaping and amplification on chip.
}
     \label{fig:soliton_applications}
\end{figure*}

Achieving ultrashort pulse generation in the mid-infrared from compact chip-scale devices may be transformative for fundamental science allowing to probe the temporal dynamics of ultrafast material processes occurring at mid-infrared frequencies --- from phonons to polaritons to interactions of complex organic molecules. The ability to generate arbitrary pulse sequences could have an impact on two-dimensional IR spectroscopy and miniaturize pump-probe systems (Fig. \ref{fig:soliton_applications}). Furthermore, the waveform-controlled QCLs may be granted an entry into nonlinear photonics as, for example, pump sources for nonlinear chalcogenide fibers for supercontinuum generation, which will open the door to octave-spanning self-referenced combs for absolute frequency metrology in the mid-infrared. From the technology perspective, mastering design and fabrication of QCLs with integrated couplers could be a steppingstone towards more complex active integrating photonic circuit architectures implemented in generic semiconductor laser material platforms, not exclusive to QCLs, but available for other material platforms, such as interband cascade lasers, quantum well and quantum dot lasers.

\subsection{New emerging active resonator platforms}

The concepts described in this review article in the context of QCL active resonators can potentially be extended to a variety of other platforms. The success in soliton generation via coherent driving in QCLs presents a blueprint that can be extended to lasers based on other III-V materials --- such as quantum well~\cite{parker2011monolithically}, quantum dash~\cite{Rosales2012HighLasers}, and quantum dot~\cite{Auth2020PassiveLasers,dong2023broadband} gain media. Such a crossover indicates a robust framework for developing new soliton-generating systems, leveraging the nuanced control of light-matter interactions inherent in these media. In contrast to the traditional ways of pulse generation via passive or active mode-locking, coherently driven solitons may offer the advantage of lower pulse timing jitter and higher efficiency because of the absence of intracavity saturable absorbers and gain modulation. 

Heterogeneous~\cite{justice2012wafer,Xiang2021LaserSilicon,marinins2022wafer,Xiang20233DPhotonics,lihachev2022low,xiang2021high,bian20233d} and monolithically~\cite{liu2015quantum,Liu2018PhotonicSilicon,shang2021perspectives,de2023gaas} integrated III-V gain media within silicon photonics are pivotal in this evolution, offering a synergistic blend where III-V semiconductors provide active lasing capabilities and silicon ensures high integration density with ultralow loss silicon nitride (SiN) components. This integration not only has the potential to enhance the efficiency of soliton generation but also provides a new playground for the exploration of the multimode coherent states of light induced by nonlinearities in both passive and active components~\cite{hermans2021high}.

Furthermore, the development of active ion-doped SiN waveguides and resonators introduces a new dimension to the field. By doping silicon nitride with active rare-earth-ions, these structures gain the ability to amplify light, paving the way for compact, integrated optical amplifiers and lasers that can support soliton dynamics~\cite{Liu2022AAmplifier,Gaafar2023FemtosecondChip,liu2023fully}. This progress is instrumental in overcoming the traditional limitations associated with passive SiN devices, offering a path toward more versatile and robust photonic circuits.

Additionally, the recent integration of titanium sapphire lasers within photonic circuits marks another promising platform for the exploration of soliton mode-locking in hybrid integrated photonic platforms~\cite{grivas2018generation,wang2023photonic}. Known for their broad tuning range and high power output, Ti:sapphire lasers incorporated into integrated photonic devices can provide a powerful and tunable source for soliton generation. This integration not only advances the field of ultrafast optics but also holds promise for a myriad of applications ranging from spectroscopy to quantum optics~\cite{morland2023hong}, signifying a step towards more sophisticated and multifunctional nonlinear photonic systems.

\bigskip

\subsection*{Acknowledgements}
We thank L. Columbo for reading this manuscript and useful discussions, and Harvard (T. Letsou), TU Wien (B. Schwarz, N. Opacak. S. Dal Cin) and other (L. A. Lugiato, F. Prati, M. Brambilla, L. Columbo) researchers for collaborations leading to several results referenced in this review. This review is based on work supported by the National Science Foundation under Grant No. ECCS-2221715.

\subsection*{References}

\bibliographystyle{naturemag_noURL}
\bibliography{references}

\end{document}